\documentstyle[12pt]{article}
\newcommand{\be}{\begin{equation}}
\newcommand{\ee}{\end{equation}}
\begin{document}
\begin{center}
{\bf On Properties of Boundaries and Electron Conductivity\\ in Mesoscopic Polycrystalline Silicon Films for Memory Devices}\\
G.P. Berman, G.D. Doolen, R. Mainieri, and J. Rehacek\\
Theoretical Division and CNLS, Los Alamos National Laboratory, Los Alamos NM 87545\\
D.K. Campbell\\
Department of Physics, University of Illinois at Urbana-Champaign,\\
1110 West Green St., Urbana, IL 61801-3080\\ 
V.A. Luchnikov\\
Institute of Chemical Kinetics and Combustion, Siberian Branch \\of Russian Academy of Sciences, Novosibirsk, 630090, Institutskay 3 Street, Russia\\ 
 K.E. Nagaev\\
Institute of Radio-Engineering and Electronics,
Russian Academy of Sciences,
Mokhovaya Street, 11, \\
103907 Moscow, Russia\\ \ \\

Abstract
\end{center}
We present the results of molecular dynamics modeling on the structural properties of grain boundaries (GB) in thin polycrystalline films. The transition from crystalline boundaries with low mismatch angle to amorphous boundaries is investigated. It is shown that the structures of the GBs satisfy a thermodynamical criterion suggested in \cite{keblinski}. The potential energy of silicon atoms is closely related with a geometrical quantity -- tetragonality of their coordination with their nearest neighbors. A crossover of the length of localization is 
observed. To analyze the crossover of the length of localization of the single-electron states and properties of conductance of the thin polycrystalline film at low temperature, we use a  two-dimensional Anderson localization model, with the random one-site electron charging energy for a single grain (dot), random non-diagonal matrix elements, and random number of connections between the neighboring grains. The results on the crossover behavior of localization length of the single-electron states and characteristic properties of conductance are presented in the region of parameters where the transition from an insulator to a conductor regimes takes place. 
\newpage
Recently it was demonstrated that thin polycrystalline silicon
films are promising materials for future room
temperature single-electron devices \cite{y1}-\cite{polySi}.
Main reasons which make this material so attractive are the following:
I). Usually, the film's thickness varies from 1 to 5 nm, and 
the average lateral grain size is 10 nm or less. In this case, 
the energy of an electron in a single grain, $E_e$,  is bigger than the 
thermal energy even at room temperature, $E_e>T=300^oK$.
II). It is believed that in these films, the characteristic
resistance, $R_b$, of the potential barriers between the grains is big 
enough, $R_b>R_c=h/e^2\sim 25k\Omega$. If both these conditions 
are satisfied, an electron is strongly localized in the grain. 
At the same time, one can regulate (up to some extent) the electron 
conductivity in these films by varying the gate voltage, and creating 
a current channel. Those electrons which are stored in the grains 
(storage dots) create a Coulomb repulsion for those electrons which 
are involved in the current channel. This allows one to implement
memory operations in these films at room temperature using a Coulomb blockade 
effect. Different implementations of these ideas have been discussed, for 
example, in \cite{y1}-\cite{polySi}. To satisfy both above mentioned  conditions, the boundaries between the nanocrystalline grains in these films  play an important role. For example, one of the most important characteristics of the electron transport in polysilicon films is connected with the distribution of crystalline and amorphous grain boundaries (GB) \cite{electron}. 
At room temperature, the main factor which determines the structure
of a GB is a mutual mis-orientation of the neighboring crystalline
grains. At present there does not exist a consistent theoretical approach for
the description of the GBs in thin polycrystalline films.\\
{\bf I}. \underbar{Molecular dynamics (MD) simulation of grain boundaries at room temperature.}

The initial configuration of the polysilicon film was generated 
using the Voronoi triangulation algorithm: first, the centers of the grains were defined; 
then each grain was obtained by filling the 
space, nearest to the center of the grain, with the diamond lattice 
at a chosen orientation. The atoms with very high potential energies were eliminated from the GBs. The film had the dimensions $16.7\times 16.7\times 3.1~nm$, and contained 44174 atoms in 12 grains (Fig. 1). The majority of the grains in our model were oriented randomly, so  they had both tilt and
twist mis-orientation. Four grains, namely the 3,4,7 and 8-th were generated
so that the mis-orientation tilt angle between the 3-rd and 4-th grains 
was $\alpha_{3-4}=11.0^{o}$; between 4-th and 7-th grains
$\alpha_{4-7}=11.0^{o}$; and between 7-th and 8-th grains, 
$\alpha_{7-8}=15.0^{o}$. The $GB_{3-4}$ and the $GB_{7-8}$ are close to
(1,1,1)-interface, and the $GB_{4-7}$ is close to (0,0,1)-interface.  
The (1,1,1)-planes in these four grains are perpendicular to the
plane of the figure. The average linear size of a grain in our model,
$d_{gr}\simeq 4.8~nm$, is of the same order as for the grains in the experimental polycrystalline silicon films obtained recently by thermal annealing of amorphous $Si$ deposited by decomposing silane on $SiO_{2}$ substrate \cite{polySi}.
Periodic boundary conditions were used in the $X$ and $Y$ directions,
and free boundary conditions were used in the $Z$ -- direction.
 It is shown that the potential energy of the atoms closely correlates with
the degree of their tetragonal coordination with the nearest neighbors.  
To estimate the quality of the tetragonal coordination of the $j$-th atom,
we calculated the value,
${\cal T}_j=\sum_{l<k}(l_{jl}-l_{jk})^{2})/15{\bar l}^2_j,$
 which is called the {\it tetragonality} \cite{tetra}.
Here $l_{jk}$  is the length of
the {\it k}-th edge of the tetrahedron of general shape
formed by the four nearest neighbors
of the $j$-th atom; $\overline{l_j}$ is the average length of the edge.
By definition, ${\cal T}_j$ is equal to zero for an ideal
tetrahedron (as it is the case of ideal diamond structure), and
increases if the shape of the tetrahedron is distorted. The value,
${\cal T}_j$, is correlated with the
dispersion of the edge lengths of the tetrahedron. 
According to the criterion suggested in \cite{tetra}, the spatial figure 
formed by four points can be recognized as having the ``good tetrahedral
shape'' if  ${\cal T}_j$ is
less than or equal to ${\cal T}^{(c)}=0.018$. 
The value of tetragonality of an atom is very sensitive
to the number of atoms in the first coordination shell, $z$.
In the bulk amorphous phase of silicon, for which $z_{a}\simeq 4.04$, the average tetragonality is equal to ${\cal T}_{am}=0.015$ which is slightly less than the critical value ${\cal T}^{(c)}$.
In liquid silicon (in which the first coordination shell  
consists in average of $z_{lq}\simeq 4.46$ neighbors), the average value of
tetragonality is, $<{\cal T}>_{lq}\simeq = 0.034$.  
In Table I, we present the data for the potential energy per atom,
density and average tetragonality, for the longest GBs (for which it
was possible to make measurements with  reasonable accuracy).    
Densities of the tilt crystalline GBs between the 3,~4,~7 and 8-th grains are
practically the same, or slightly smaller, than the density of the bulk
crystal, $\rho_{cr}$. The densities of other GBs are 
systematically larger than $\rho_{cr}$. In covalent materials
with friable structure, such as silicon and germanium, some 
GBs, indeed, contract. This was observed in X-ray
diffraction experiments for (1,0,0) twist GBs in Ge \cite{Ge}
and in MD simulation of the (1,1,1)-twist boundaries in
silicon \cite{twist111}. It is remarkable that the contraction is
observed for the twist boundaries. In such GBs, the covalent
bonds are most probably stretched rather than squeezed, and the system
demonstrates a tendency to restore their lengths at the cost of a reduction
of the GB's volume. MD simulations show that in our model of polysilicon film, the majority of the GBs between the grains with random
mutual orientation are disordered.
The grains with small mutual mis-orientation are connected
by crystalline GBs. The GBs with ``medium-angle'' mis-orientation have a complex inhomogeneous structure. These boundaries consist of crystalline connections interspersed with disordered regions. In
disordered regions, the misfit between the crystal structures of the neighboring grains is compensated by
non-crystalline arrangements of atoms, such as 5- and 7- fold rings.
In general, the structure of the GBs satisfies the thermodynamical criterion
suggested in \cite{keblinski}.
The potential energy of the silicon atoms closely correlates 
with the tetragonality, ${\cal T}$, of their coordination with their nearest neighbors.  
The majority of atoms in the polysilicon grain boundary are well
coordinated tetrahedrally, even if they are arranged in a non-crystalline manner in
the high-angle GB. From the observed small values of the average tetragonality, it follows that the high-angle disordered GBs can
be characterized as amorphous.

The results obtained are important for better understanding of structural
properties of thin  polycrystalline silicon films, which 
have been used recently for memory devices. However, for obtaining more
reliable data, larger polysilicon simulations should be considered.\\
{\bf II.} \underbar{Crossover behavior of the localization length of single-electron states,}\\
\underbar{ and the properties of conductivity.}  

To model the properties of conductive electrons in thin polycrystalline
silicon
film at low  temperature, we use an approach  based on the two-dimensional
tight-binding Anderson model. Each nanosilicon grain (with the coordinates
$(m,n)$ in $(X,Y)$-directions) is considered as an individual dot. The
characteristic energy of quantization in the $Z$-direction,
$E^{z}$, for the lowest level, is of the order, $E^{z}\sim 1-300$ meV
\cite{y1,y2}.
We assume that the total one-site
electron energy is $E_{m,n}=<E>+\delta E_{m,n}$, where $\delta E_{m,n}$ is a
random variable (as the thickness of the film fluctuates).
At low temperature, the electron conduction is governed by
quantum tunneling between the dots, which we describe by the matrix elements,
$V_{m,n}^{(m^\prime,n^\prime)}$. Below we consider only the simplest
 case of the neighboring transitions:  $V_{m,n}^{(m\pm 1,n\pm 1)}=
<V>+\delta V_{m,n}$, where $<V>$ is the average value of the off-diagonal
matrix
elements, and $\delta V_{m,n}$ describes the  fluctuations related with the
random characteristics of the potential barriers between the neighboring dots.
Let us now estimate the  number of conduction electrons which occupy an
individual dot. The maximum value of a 2D electron density in the film can be
chosen as: $n_e\sim 10^{11}-10^{12}cm^{-2}$. Consider a subsystem of
conduction
electrons in a polysilicon film as a two-dimensional one, with the size of the
active region 100 nm$\times$100 nm \cite{y1,y2}. Then, the number of
conduction
electrons, ${\cal N}$, in this region is of the order ${\cal N}=10-100$.
If the
average size of the grain (dot) in $(X,Y)$ plane is of the order 10nm,
then one
has, on average, 100 dots in the active region. So, the average number of
conductive electrons in a dot is of the order, 0.1-1. In this paper, we
consider
the conduction electrons as noninteracting.
To describe the electron subsystem at low temperature, we use a 2D Anderson
tight-binding model with the Hamiltonian
${\cal H}=\sum_lE_l|l><l|+\sum_{l,k}V_{l,k}|l><k|$, where
$E_l$ describes the diagonal disorder.
Fig. 2 shows the dependences of the average length of localization, $<L>$, and
the dispersion of the length of localization, $D(L)=<(L-<L>)^2>$, of the
single-electron states, $\Psi^{(i)}(m,n)$, as a function of the average
value of
the matrix element $<V>$. (The size of the lattice is ($M,N$)=($18,12$). Zero
boundary conditions were chosen.)  In Fig. 2, $<E>=50$ meV; $E_{m,n}\in
[50-12.5;50+12.5]$ meV; $\delta V_{m,n}\in [0.9;1.1]<V>$ meV. For each state,
$i=1,...,MN$, the length of localization was introduced as
$L^{(i)}=(L^{(i)}_m+L^{(i)}_n)/2$. We calculated $L^{(i)}_m$  we as
$L^{(i)}_m=\sqrt{\sum_{m,n=1}^{M,N}(m-\bar m^{(i)})^2|\Psi(m,n)|^2}$, where
$\bar m^{(i)}=\sum_{m,n=1}^{M,N}m|\Psi(m,n)|^2$. (Similar expressions
were used
to calculate  $L^{(i)}_n$.) As one can see from Fig. 2, the fluctuations
of the
average localization length, $D(L)$, exhibit a characteristic maximum at
$<V>\approx <V_c>=3.5$ meV.  For $<V><<V_c>$, the system is essentially
insulating. For $<V>><V_c>$, the system exhibits  metallic
properties. We are presently investigating this crossover behavior in
connection
with the metal-insulator transition (MIT). Finally, in Fig. 3, we present the
characteristic results of numerical simulations of the dimensionless
conductance in this system, $\sigma=\sum |T_{\alpha,\beta}|^2$, where
$T_{\alpha,\beta}$ is the transmission amplitude from channel $\alpha$ to
channel $\beta$.  The method we used is based on a Green's functions approach
\cite{bn}.
The lattice size is (11,11), with 11 channels in the left and the right
electrodes. In Fig. 3, the horizontal axis is the dimensionless value of the
diagonal disorder, $\delta E$. The dimensionless off-diagonal matrix
element was chosen equal to
$V=1$ in the active region and in the electrodes. The results shown in
Fig. 3 correspond to the dimensionless Fermi energy $E_F=0$.
In Fig. 3 the data
indicated by $\diamond$ (a) corresponds to the randomization of only 
two
boundaries, ($n=1$, $m=1,...,M$) and ($n=N$, $m=1,...,M$). The data indicated
by $+$ (b) corresponds to the randomization of the whole active region. One can
see that increasing the diagonal disorder leads to a crossover behavior of the
conductance in the case when only the boundaries are randomized. The results presented in this paper can be useful for performance the memory devices based on thin polycrystalline silicon films. Our further work is related to studying the influence of the interfaces $Si/SiO_2$ on the conductivity in polycrystalline silicon films, and on the noise component.  

We would like to thank D.K. Ferry and P. Lomdahl for useful
discussions. V.A.L. and K.E.N. thank the Center for Nonlinear Studies, Los
Alamos National Laboratory, for hospitality.
 This work was supported by the Linkage Grant 93-1602
from the NATO Special Programme Panel on Nanotechnology, and by the Defense
Advanced Research Projects Agency.
\newpage
\newpage 
{
$$
\begin{tabular}{|c|c|c|c|} \hline
$ ~ $ & $U~(eV/at), \pm 0.04$ & $\rho~(g/cm^3), \pm 0.02$ & $<{\cal T}>, \pm 0.0003$ \\ \hline

$GB_{3-4}$ &   $-4.24$   &  $2.29$  &  $0.005$\\ \hline
$GB_{4-7}$ &   $-4.18$   &  $2.31$  &  $0.014$\\ \hline
$GB_{7-8}$ &   $-4.20$   &  $2.34$  &  $0.011$\\ \hline
$GB_{3-6}$ &   $-4.19$   &  $2.39$  &  $0.010$\\ \hline
$GB_{6-7}$ &   $-4.16$   &  $2.39$  &  $0.013$\\ \hline
$GB_{4-1}$ &   $-4.17$   &  $2.41$  &  $0.013$\\ \hline
$GB_{1-5}$ &   $-4.18$   &  $2.38$  &  $0.011$\\ \hline 
$GB_{5-2}$ &   $-4.09$   &  $2.36$  &  $0.017$\\ \hline
$GB_{9-10}$&   $-4.21$   &  $2.37$  &  $0.012$\\ \hline
$GB_{10-11}$&  $-4.15$   &  $2.38$  &  $0.015$\\ \hline
$GB_{9-12}$&   $-4.14$   &  $2.38$  &  $0.017$\\ \hline
$a-Si$     &   $-4.15^*$ &  $2.29^*$&  $0.015^*$\\ \hline
$cr-Si$    &   $-4.335$  &  $2.324$ &  $0$\\ \hline
\end{tabular}
$$
{\it Table~I:}\quad Potential energy,
density and average tetragonality of grain boundaries
in the polysilicon model. $^*$Data for a-Si obtained from the model \cite{andreas}

\newpage

\begin{center}
{\bf Figure Captions}
\end{center}
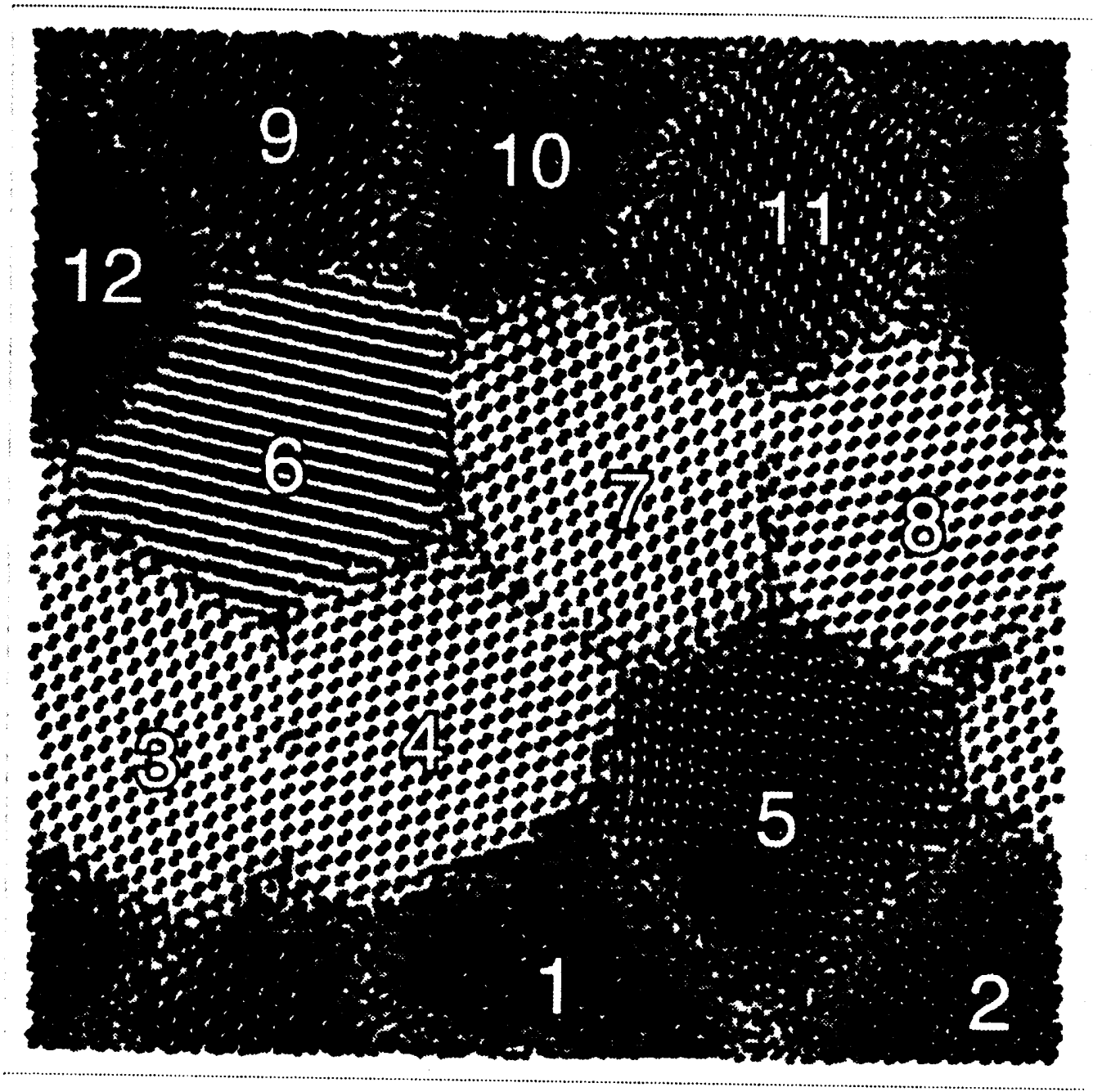
Fig. 1.\quad The $(X,Y)$-projection of the polycrystalline film.\\ \ \\
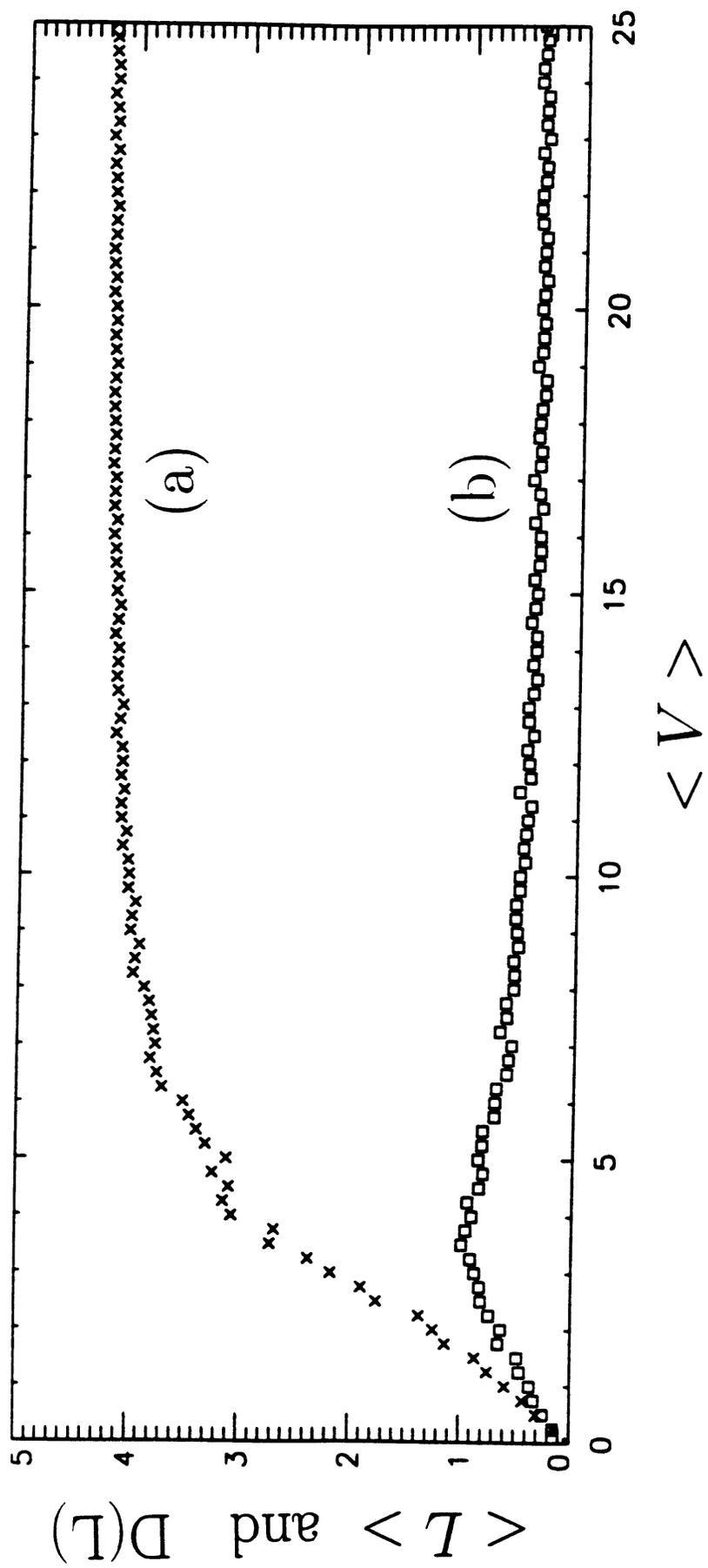
Fig. 2.\quad  Demonstration of the crossover behavior of the average localization length of the single-electron states. Dependence of the average localization length $<L>$ (a), and of the dispersion $D(L)$ (b) on  the average value of the matrix element, $<V>$. $(M,N)=(18,12)$; $<E>=50$ meV; $\delta E=25$ meV; $V=\in [0.9;1.1]<V>$ meV.\\ \ \\
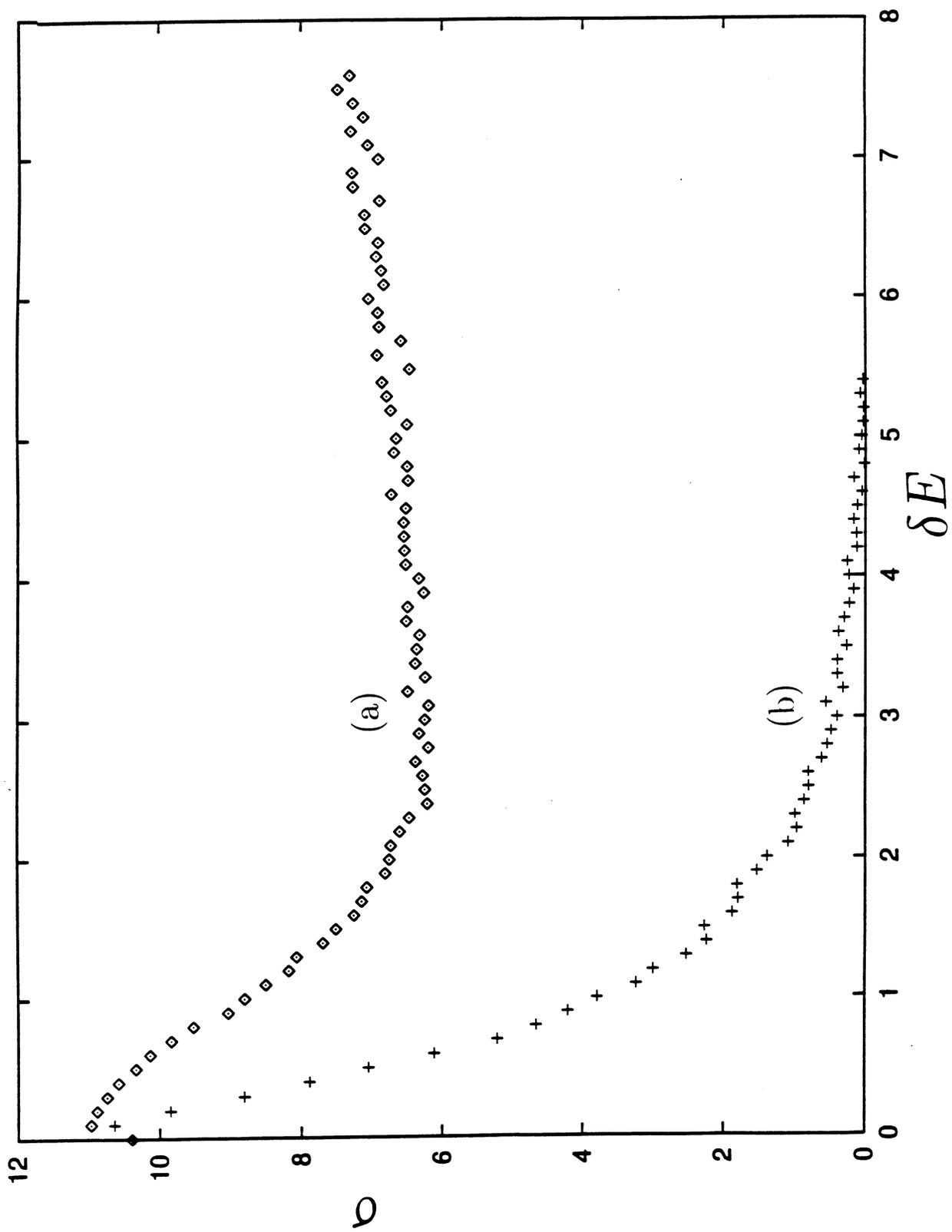
Fig. 3. \quad Dependence of the conductance, $\sum T_{\alpha,\beta}$, on the disorder parameter $\delta E$. The lattice size $(M,N)=(11,11)$. 11 channels in the left and  right electrodes were used. Dimensionless parameters: $V=1$; $E_F=0$; (a) Two of
the boundaries are randomized: ($n=1$, $m=1,...,M$), and ($n=N$, $m=1,...,M$);
(b) The whole active region is randomized.

\end{document}